\documentclass[aip,
 amsmath,amssymb,
 reprint,
]{revtex4-1}
\usepackage{graphicx}
\usepackage{dcolumn}
\usepackage{bm}
\usepackage{amsmath}
\usepackage[utf8]{inputenc}
\usepackage[T1]{fontenc}
\usepackage{mathptmx}
\usepackage{etoolbox}
\usepackage{xr}
\usepackage{hyperref}
\newcommand{\baln}{B$_{x}$Al$_{1-x}$N }

\newcommand{\balnn}{B$_{x}$Al$_{1-x}$N}
\makeatletter
\def\@email#1#2{%
 \endgroup
 \patchcmd{\titleblock@produce}
 {\frontmatter@RRAPformat}
 {\frontmatter@RRAPformat{\produce@RRAP{*#1\href{mailto:#2}{#2}}}\frontmatter@RRAPformat}
 {}{}
}%
\makeatother
\begin{document}
\preprint{AIP/123-QED}
\title{First-principles band alignment engineering in polar and nonpolar orientations for wurtzite AlN, GaN, and B$_x$Al$_{1-x}$N alloys}
\author{Cody L. Milne}
\affiliation{Department of Physics, Arizona State University}
\author{Arunima K. Singh}
\affiliation{Department of Physics, Arizona State University}
\email{arunimasingh@asu.edu}
\date{\today}

\begin{abstract}
Boron aluminum nitride (B$_x$Al$_{1-x}$N) is a promising material for next-generation electronic and optoelectronic devices due to its ultra-wide bandgap, high thermal stability, and compatibility with other III-nitride semiconductors. Despite its potential, the band alignments of B$_x$Al$_{1-x}$N remain largely unexplored, although this information is essential for device design. In this study, we compute the valence and conduction band alignments of nonpolar ($a$-plane) and polar ($c$-plane) \balnn, and compare them with those of AlN and GaN. Using density functional theory, many-body perturbation theory, $GW_0$ method, and a novel passivation scheme, we find that they have near-zero valence band alignments for low-$x$ \balnn/AlN, while higher compositions ($x > $0.333) exhibit type I or II band alignments. The band alignments also show a notable dependence on surface polarity and the tetrahedral distortion of the \baln structures. Our computed offsets are in good agreement with available experimental data. Due to their low valence band alignments and higher conduction band alignments, the \balnn/AlN heterostructures could be well suited for high-electron-mobility transistors and ultraviolet light-emitting diodes. The band alignments of B$_x$Al$_{1-x}$N determined in this study provide essential design guidelines for integrating these ultra-wide bandgap alloys into advanced semiconductor technologies. 
\end{abstract}

\maketitle
\maketitle
Boron aluminum nitride (\balnn) is a ternary group-III nitride material that has seen increasing interest in recent years due to its array of intriguing properties. In the wurtzite phase (w-), its tunable ultra-wide bandgap (6.2-7.4 eV) extends the bandgap range available to III-nitride electronic materials, making it particularly attractive for advanced electronic and optoelectronic devices in the deep-ultraviolet range\cite{Milne2023, Milne2024}. Its excellent thermal conductivity and high chemical stability mean it could be used in next-generation high-power and high-frequency devices\cite{Kudrawiec2020,Milne2023,Milne2024}. In addition to these excellent properties, it has recently been demonstrated that as little as 2~\% of boron incorporation into AlN yields ferroelectric switching capabilities\cite{Hayden2021, Zhu2021, Calderon2023, Milne2024}, and it has also been shown to exhibit a dielectric constant much higher than AlN or BN, e.g. in B$_{0.07}$Al$_{0.93}$N a value of $\sim 16\varepsilon_0$ is predicted, compared to $\sim 9.7\varepsilon_0$ in AlN and $7.3\varepsilon_0$ in w-BN\cite{Hayden2021, Milne2024, Savant2025}. These properties could enable the use of \baln for novel dielectric or ferroelectric layers\cite{Hayden2021,Milne2024,Savant2025}, and thus \baln has a huge potential in III-nitride compatible heterojunction-based devices. 

Despite growing interest in \baln for device applications, systematic studies of its electronic properties relevant to heterojunctions are still scarce. 

Experimental investigation of such heterojunctions remains challenging due to the difficulty in obtaining high-purity, phase-pure, and high-crystallinity samples of \balnn\cite{Akasaka2007, Li2015, Li2017, Sun2017, Sun2018a, Sun2018b, Tran2020, Vuong2020, Hayden2021, Sarker2020, ZhangQ2022, Zhu2021, Wolff2021, Calderon2023}. Additionally, theoretical studies of certain surfaces in wurtzite group-III-nitrides, such as the (0001) $c$-plane, are complicated due to the electric field generated by the polarization-induced charge displacement\cite{Yoo2021}, which further limits the understanding of the fundamental factors that govern the properties of the heterojunctions. 

\begin{figure*}
 \includegraphics[width=2\columnwidth]{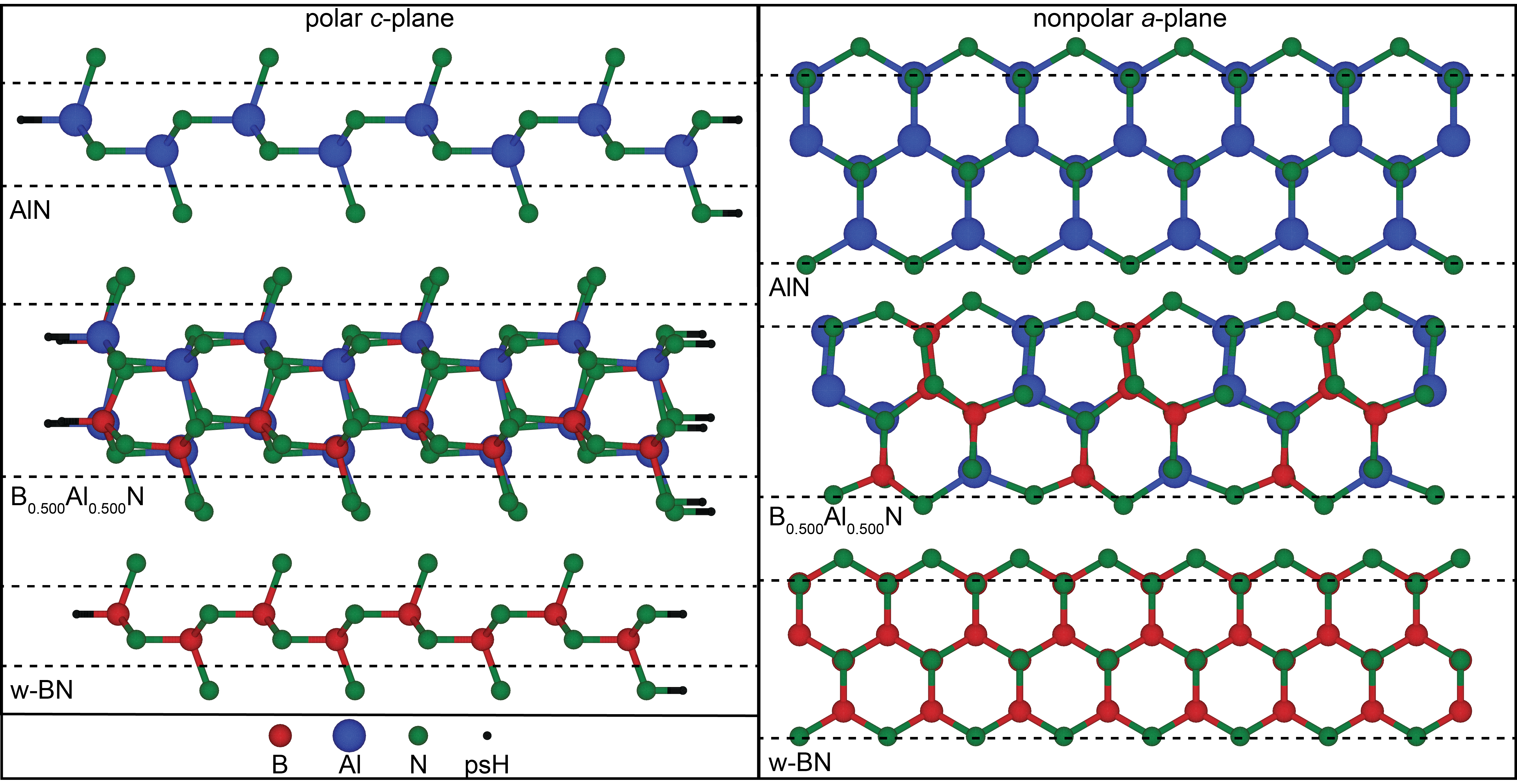}
 \caption{Slab structures in the $c$-plane (left) and $a$-plane (right) orientations for AlN, B$_{0.500}$Al$_{0.500}$N, and w-BN. The slab structures are fully passivated by pseudohydrogen (psH) atoms in the $c$-plane. The $a$-plane orientations are unpassivated. The dashed lines indicate the slab lattice with surface normals in the horizontal direction. Red, blue, green, and black spheres represent boron, aluminum, nitrogen, and psH atoms, respectively.}
 \label{fig:structure_main_text}
\end{figure*}

Among various heterojunction properties, the band alignments at heterojunction interfaces critically determine carrier confinement, transport, and recombination behavior; thus, understanding these alignments in \balnn-based systems is essential for device design. However, only a few investigations have examined the band alignments of \baln to date.

Experimentally, Sun \emph{et al.} measured the band alignments in B$_{0.14}$Al$_{0.86}$N thin films with respect to AlN, GaN\cite{Sun2018a, Sun2018b}, and Al$_{0.7}$Ga$_{0.3}$N\cite{Sun2017}. Rather \emph{et al.} measured the band alignments for B$_{0.13}$Al$_{0.87}$N/GaN and B$_{0.13}$Al$_{0.87}$N/AlN heterojunctions\cite{Rather2023}. Only Ota \emph{et al.} has published a theoretical study on band alignments in \baln using the empirical solid-state energy (SSE) approach, which does not take into account effects of surfaces\cite{Ota2022} which have significant impact for polar materials and interfaces\cite{BiswasRice2025, Milne2025}. 

In this work, we make the first surface-dependent predictions of \baln band alignment in the complete $x=0$ to $1$ range using first-principles simulations. 

For determination of the band alignment alignment, we follow the two-step process where in the first step, the valence-band maximum (VBM) and conduction-band minimum (CBM) energies are computed for each bulk structure.\cite{Tsai2020} Alongside 17 ground-state structures of \balnn\cite{Milne2023, Milne2024}, we also investigate the band alignment of GaN, AlN, and w-BN. In a previous work, we determined the 17 ground-state structural configurations of \baln using DFT and the cluster expansion formalism\cite{Milne2023}. In this work, we also introduce $x=0.083$ and $x=0.125$ \baln models to investigate the low-$x$ regime, with their structures taken to be the lowest formation energy structures from three random alloys. The VBM and CBM energies in the bulk structure are referenced to the macroscopic average potential in the bulk structure, $\bar{V}_\mathrm{macroscopic}^\mathrm{bulk}$. In the second step, we simulate a finite-slab of the material to align $\bar{V}_\mathrm{macroscopic}^\mathrm{slab}$ with the vacuum energy, $E_\mathrm{vac}$, establishing a universal reference energy that allows comparison between different materials. Thus this two-step process leads to computation of a surface-dependent VBM energy, $E_\mathrm{VBM}^\mathrm{slab}$, as

\begin{equation} \label{eq:potential}
  E_\mathrm{VBM}^\mathrm{slab} = (E_\mathrm{VBM}^\mathrm{bulk} - \bar{V}_\mathrm{macroscopic}^\mathrm{bulk}) + \bar{V}_\mathrm{macroscopic}^\mathrm{slab} - E_\mathrm{vac}.
\end{equation}
Similarly, the CBM energy, $E_\mathrm{CBM}^\mathrm{bulk}$, can also be obtained. 
\autoref{fig:structure_main_text} shows a schematic of some slab models of the $c$-plane and $a$-plane for AlN, B$_{0.500}$Al$_{0.500}$N, and w-BN, respectively. In total, 104 distinct surfaces were studied in this work. 

This two-step approach has long been used for computing band alignment of non-polar surfaces, even in polar materials.\cite{Moses2011, Hinuma2014, Hinuma2017, Tsai2020} However, the spontaneous polarization, $P_s$, induced electric fields interfere with the potential alignment, making this approach unsuitable for polar surfaces such as wurtzite $c$-plane III-nitrides and \balnn. We employ a recently developed passivation scheme by Yoo \emph{et al.} \cite{Yoo2021} to consider the electric fields due to spontaneous polarization in finite slabs, addressing the long-standing challenge of computing band alignments in materials with spontaneous polarization. The passivation involves adsorbing pseudohydrogen atoms (psH) with $P_s$-dependent modified valency to one surface; unlike "conventional passivation scheme", which only uses psH with valency of the missing bond(s)\cite{Shiraishi1990}. Further information on the passivation scheme is available in the Supplementary Information Figure S1 and S2. In the Supplementary Information, the DFT-computed $P_s$ with respect to the zincblende structure, $P_s^{\mathrm{zb}}$, and the modified psH for all the polar slabs are presented in detail. In addition, information on DFT computed $P_s^\mathrm{hex}$ and piezoelectric constants is available in the Supplementary Information. We emphasize here that this work presents the first DFT-computed $P_s^{\mathrm{zb}}$ in \baln in Figures S3 in the Supplementary Information.

All \emph{ab-initio} simulations reported in this study were performed using the Vienna Ab Initio Simulation Package (VASP) \cite{Kresse1, Kresse2, Kresse3, Kresse4} package using the Projector-Augmented-Wave (PAW) formalism\cite{Perdew20} with the PBE \cite{Blochl1994, Perdew19, Perdew20} exchange-correlation functional for all calculations. Slab calculations were performed without ionic relaxation, therefore predicted band alignments do not include the effects of strain due to heterojunction lattice mismatch, interfacial relaxation, or dependence on experimental growth conditions. A dipole correction\cite{Freysoldt2020} was implemented along the $z$-axis to correct the spurious electric fields in the vacuum regions. For the slab calculations, the total energy was converged within $10^{-7}$ eV using $\Gamma$-centered $k$-grids of 1000 $k$-points per \AA$^3$. An energy cutoff of 520 eV was used for all calculations. Thicknesses of 8 bilayers were used for all $c$-plane ($15.3-19.0$ \AA) slabs, which we have determined to be below the threshold for electric breakdown by analyzing the thickness-dependent electric fields. For nonpolar $a$-plane slabs, a thicknesses of at least 13 layers ($\geq20.0$\AA) was employed. These parameters allow convergence of $\bar{V}_\mathrm{macroscopic}^\mathrm{slab} - E_\mathrm{vac}$ to within 0.01 eV. 

Many unique slab configurations can often be constructed from the bulk \baln structures due to the varying fraction of atoms in the lattice. Therefore, we calculate band alignments for every slab for each $x$ generated using \texttt{pymatgen}\cite{Ong2013}, which generates slab models by rotating the bulk structure into the desired orientation and cleaving it to generate all unique surface terminations for a given thickness. We obtain $\bar{V}_\mathrm{macroscopic}^\mathrm{slab} - E_\mathrm{vac}$ for each slab model along with $E_\mathrm{VBM}^\mathrm{bulk} - \bar{V}_\mathrm{macroscopic}^\mathrm{bulk}$ from corresponding bulk calculations to calculate $E_\mathrm{VBM}^\mathrm{slab}$ and $E_\mathrm{CBM}^\mathrm{slab}$ for all slabs of the same $x$ and surface plane, with error bars indicating standard deviations. 

$E_\mathrm{VBM}^\mathrm{bulk}$, $E_\mathrm{CBM}^\mathrm{bulk}$, and other electronic properties of bulk AlN, w-BN, and ground-state bulk \baln configurations were computed in our previous work\cite{Milne2024}. The mid-gap approximation was employed to obtain the $GW_0$ $E_\mathrm{VBM}^\mathrm{bulk}$ and $E_\mathrm{CBM}^\mathrm{bulk}$ from the PBE values \cite{Liang2013}. Given the severe band-gap underestimation in PBE\cite{Biswas2023}, we used $GW_0$ quasiparticle band gaps\cite{Biswas2021,Biswas2023,Milne2024,Biswas2025}, which give excellent agreement with experimental values and are available in the Supplementary Information Figure S4 and Table S1. 

\begin{figure}[h!]
 \includegraphics[width=0.85\columnwidth]{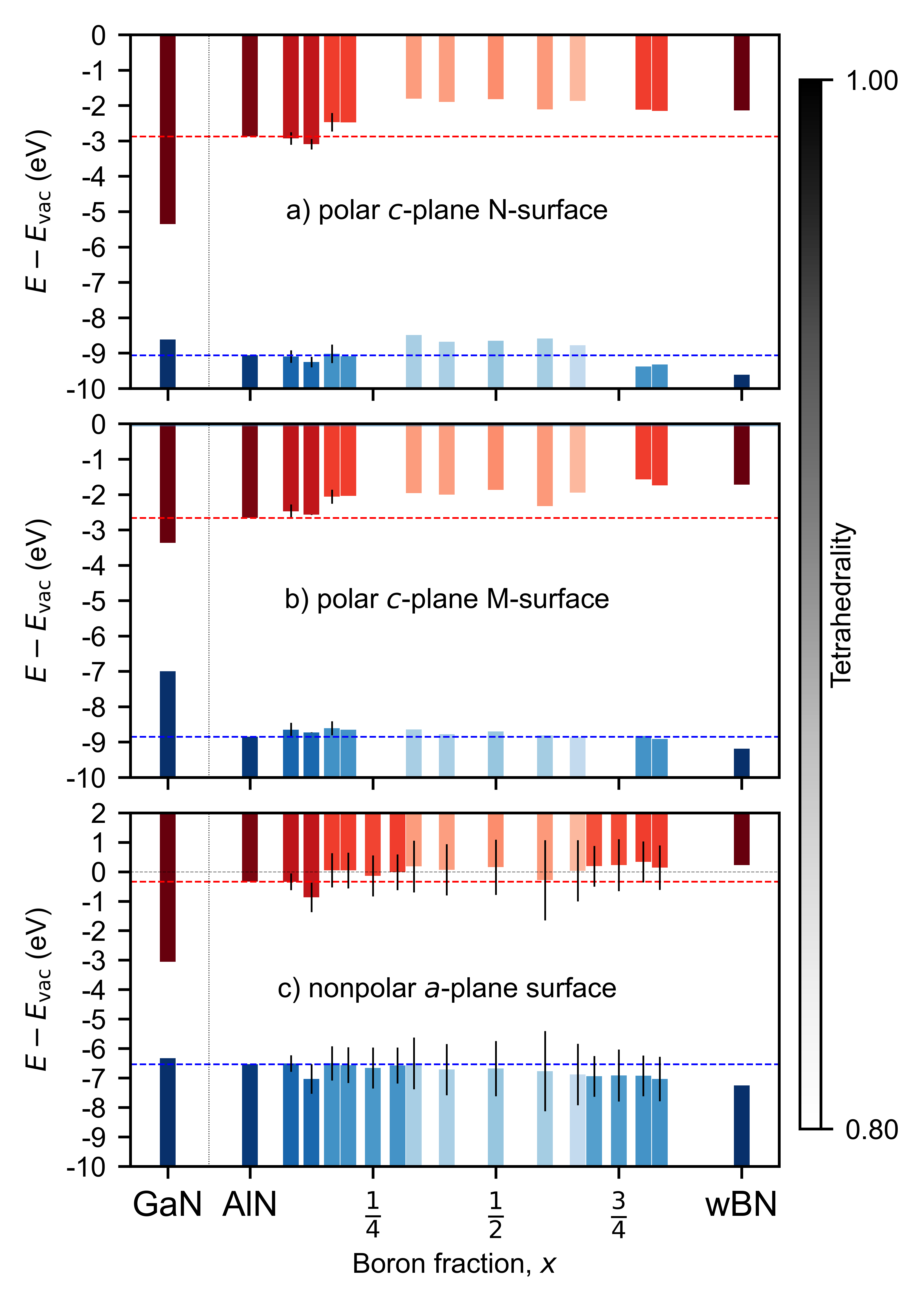}
 \caption{Band alignments for GaN and \balnn, where $E_\mathrm{VBM}$ and $E_\mathrm{CBM}$ are represented by blue and red columns, respectively. The columns are shaded according to the average tetrahedrality in the bulk \baln structure, which are shown by the colormap. Blue and red dashed horizontal lines show the AlN $E_\mathrm{VBM}$ and $E_\mathrm{CBM}$, respectively. The vertical dashed gray lines separate the GaN and \baln results. a) Shows the $c$-plane N-surface slabs, b) the $c$-plane M-surface slabs, and c) the $a$-plane symmetric slabs. Error bars indicate standard deviations of the band alignments of the structures of the same composition.}
 \label{fig:band_alignment_no_SSE}
\end{figure}

The results of our band alignment study are shown comprehensively in \autoref{fig:band_alignment_no_SSE}, where a) and b) show the band alignments for the $c$-plane N- and M-surfaces (where M=B, Al, or Ga), and c) shows the band alignments for the $a$-plane surfaces. By taking the difference between the $E_\mathrm{vac}$-aligned $E_\mathrm{VBM}^{\mathrm{slab}}$ or $E_\mathrm{CBM}^{\mathrm{slab}}$ (hereafter referred to as $E_\mathrm{VBM}$ and $E_\mathrm{CBM}$) of different materials, we obtain the valence band offsets (VBO) or conduction band offsets (CBO) using an Anderson's model type approach\cite{Anderson1960, Tsai2020}. All $c$-plane slab models (Figures S5-S9) and their potential profiles (Figure S10-12) are available in the Supplementary Information.

Ideal interfaces between two $c$-plane III-nitrides consist of a shared nitrogen layer. Therefore we compute band alignments for $c$-plane N-surface slabs of each III-nitride to model each side of the interface. III-nitride heterointerfaces are used in a variety of device structures, e.g. AlN/Al$_x$Ga$_{1-x}$N and GaN/In$_x$Ga$_{1-x}$N heterojunctions\cite{Nardelli1997,Nardelli1997PRB}. Additionally, N-polar surfaces are recommended for Ohmic contacts due to lower barrier heights\cite{Reddy2014}. Therefore, a detailed investigation of the N-surface band alignments is necessary, the results of which are shown in \autoref{fig:band_alignment_no_SSE}a for GaN and \baln where $E-E_\mathrm{vac}$, i.e. $E_\mathrm{VBM}^{\mathrm{slab}}$ or $E_\mathrm{CBM}^{\mathrm{slab}}$, is the energy of the band edges with respect to the vacuum energy. 

To benchmark our calculations, we consider the well-studied AlN/GaN heterointerface. Prior studies agree that $E_\mathrm{VBM}$ of GaN is above that of AlN, indicating a type-I band alignment, tabulated in \autoref{tab:band_offsets}. However, there is some dependence of the VBO on the lattice strain\cite{Nardelli1997PRB} and the experimental growth temperatures\cite{King1998}. We obtain a VBO of 0.44 eV for AlN/GaN, which is in excellent agreement with the previously reported computed values of 0.44 eV\cite{Nardelli1997PRB}, as well as with XPS measurements of $0.5\pm0.3$ eV\cite{Baur1994,King1998}. The agreement can be attributed to the fact that contributions from lattice mismatch and interfacial atomic relaxation are small in AlN/GaN $c$-plane interfaces\cite{Nardelli1997}. Similarly, these contributions are also expected to be small in low-$x$ \balnn/AlN interfaces due to the low lattice mismatch\cite{Milne2023}.

The common-anion rule states that VBO are small between materials with common anions since the valence wavefunctions are typically dominated by anionic contributions in group III-nitrides\cite{Franciosi1993,Ota2020}. Similar to AlN/GaN, we observe that $E_{\mathrm{VBM}}$ in \baln and AlN are nearly equal, resulting in small VBO (\autoref{fig:band_alignment_no_SSE}a). However, there are variances with respect to $x$. At low $x$ ($x<0.333$), $E_{\mathrm{VBM}}$ is similar to that of AlN with near-0 VBO. The low lattice distortion, low lattice mismatch, and similar bandgaps in low-$x$ \baln results in similar $E_\mathrm{VBM}$ and $E_\mathrm{CBM}$ to AlN. In intermediate $x$ ($0.333 \leq x \leq 0.667$) \balnn, $E_\mathrm{VBM}$ rises above that of AlN. Here, we find a correlation between the average tetrahedrality in the \baln structures and $E_\mathrm{VBM}$. The tetrahedrality is a measure of the bonding environment of an atomic site, where 0 and 1 signify a non-tetrahedral and perfect tetrahedral bonding environment, respectively. Therefore, reduced tetrahedrality means the bonds deviate from that of ideal wurtzite bonding\cite{Milne2023, Milne2024}. The average tetrahedrality is lowest in the intermediate $x$ alloys\cite{Milne2023}, resulting in elevated $E_\mathrm{VBM}$ and $E_\mathrm{CBM}$. However, $E_\mathrm{VBM}$ becomes lower than that of AlN for high-$x$ ($x> 0.667$). These lower $E_\mathrm{VBM}$ values are due to high boron content and the low lattice distortion (due to high tetrahedrality)\cite{Milne2024}. Therefore, \balnn/AlN and \balnn/GaN interfaces could provide tunable band alignments through modulation of the boron fraction in \balnn, particularly if higher $x$ ($x \geq 0.3$) \baln is realized experimentally.

In spite of the growth challenges with \balnn, there are a few existing experimental measurements of the band alignments in \balnn/GaN and \balnn/AlN heterojunctions (\autoref{tab:band_offsets}). Measurements of B$_{0.13}$Al$_{0.87}$N/GaN\cite{Rather2023} and B$_{0.14}$Al$_{0.86}$N/GaN\cite{Sun2018b} heterojunctions have shown major disagreement at nearly the same $x$, with reported VBOs of $1.0\pm0.1 $ eV and $0.2\pm0.2$ eV, respectively. The calculated VBOs for the similar $x$ B$_{0.125}$Al$_{0.875}$N/GaN and B$_{0.167}$Al$_{0.833}$N/GaN interfaces are $ 0.63 \pm 0.15$ and $0.40 \pm 0.26$ eV respectively. Notably, the average B$_{0.125}$Al$_{0.875}$N/GaN VBO ($0.63 \pm 0.15$ eV) compares very favorably to the average VBO reported between the two experimental studies ($0.6$ eV). Thus, our calculation scheme provides a useful prediction of the experimental VBO in $c$-plane \balnn/GaN interfaces. Moreover, the discrepancy observed in calculations and experimental measurements for similar $x$ \balnn/GaN heterojunctions are likely due to differing interfacial stoichiometry near the interface. The effect of interfacial stoichiometry has been studied in Al$_x$Ga$_{1-x}$N and In$_x$Ga$_{1-x}$N\cite{Tsai2020}. 

Further study of our $c$-plane slab models revealed that a higher B composition near the surface leads to lower (more negative) $E_{\mathrm{VBM}}$ and $E_\mathrm{CBM}$. Intuitively, this is more similar to the high-$x$ \baln $E_\mathrm{VBM}$ and $E_\mathrm{CBM}$.

\begin{table*}[]
  \centering
  \begin{tabular}{c|c|c|c|c} 
    $c$-plane interface & & VBO (eV) & VBO (eV) & VBO (eV) \\\hline
    AlN/GaN & This work & $0.44$ & & \\
        & Ref. Theory & $0.44-0.73$\cite{Nardelli1997PRB}$^a$ & $0.30$\cite{Tsai2020}$^b$ & $0.28 - 0.71$\cite{He2022}$^a$ \\
         & Ref. Exp. & $0.5\pm0.3$\cite{King1998} & $0.5$ \cite{Baur1994} & $0.70\pm0.24$\cite{Martin1996} \\ \hline
    \balnn/GaN & This work & $0.48 \pm 0.18$ ($x=0.083$)& $ 0.63 \pm 0.15$ ($x=0.125$) & $0.40 \pm 0.26$ ($x=0.167$) \\
    & Ref. Exp. & & $1.0 \pm 0.1$\cite{Rather2023} ($x=0.13$) &$0.2 \pm 0.2$\cite{Sun2018b} ($x=0.14$) \\ \hline
    \balnn/AlN & This work & $0.04 \pm 0.18$ ($x=0.083$) & $0.19 \pm 0.15$ ($x=0.125$) & $-0.04 \pm 0.26$ ($x=0.167$) \\
    & Ref. Exp. & & $0.5 \pm 0.1$\cite{Rather2023} ($x=0.13$)& \\ \hline
  \end{tabular}
  \caption{The relative valence and conduction band alignments for III-nitride heterostructures calculated using the N-surface $c$-plane band alignments and their comparison to existing experimental and theoretical results in the literature.\\
  $^a$ DFT study, 
  $^b$ Hybrid functional study}
  \label{tab:band_offsets}
\end{table*}

The III-nitride M-surfaces will have different electronic properties than N-surfaces due to electronegativity differences in the surface atoms. Therefore, an understanding of how the M-surface band alignments differ is needed. For example, heterointerfaces of materials with shared group-III cations (i.e., AlN/Al$_2$O$_3$)\cite{Zoino2023} could be understood through analysis of the M-surface band alignments. Therefore, we report the $c$-plane M-surface band alignments for GaN, AlN, w-BN, and \baln in \autoref{fig:band_alignment_no_SSE}b.

In stark contrast to the GaN N-surface, the GaN M-surface $E_\mathrm{VBM}$ and $E_\mathrm{CBM}$ are dramatically higher than the GaN N-surface $E_\mathrm{VBM}$ and $E_\mathrm{CBM}$ by 1.62 eV. This large difference ($>1.4$ eV\cite{Yang2003} and $2$ eV\cite{Reddy2014}) between Ga-surface and N-surface electron affinities has been experimentally observed in GaN. As a result, the predicted AlN/GaN M-surface VBO is 1.85 eV, much higher than the 0.44 eV calculated for the N-surface structures. Interestingly, many experimental measurements of the AlN/GaN VBO have found higher values\cite{Martin1996,King1998,Li2014,He2022}, for example, a VBO of $1.36\pm0.07$ eV was measured by Waldrop \emph{et al.}\cite{Waldrop1996}. The higher measured VBO could be observed in AlN/GaN heterostructures as a result of interfacial Al/Ga mixing near the $c$-plane interface rather than an abrupt interface consisting of a single nitrogen layer. Thus, while \balnn/AlN heterojunctions have similar VBO for N- and M-surfaces, this large difference between GaN M- and N-surface $E_\mathrm{VBM}$ and $E_\mathrm{CBM}$ provides a route for band alignment tuning in \balnn/GaN heterojunctions by modulation of the interfacial stoichiometry.

Overall, the M-surface $E_\mathrm{VBM}$ trends in \baln are similar to that of the N-surface $E_\mathrm{VBM}$ in \balnn, and nearly constant with $x$. This suggests that the B or Al composition near the M-surface, as well as the tetrahedrality of the structure, has little impact on $E_\mathrm{VBM}$. However, we find that the w-BN $E_\mathrm{VBM}$ is below that of AlN by 0.34 eV. 

Construction of non-polar III-nitride heterojunctions could be useful through the elimination of the piezoelectric polarization effects\cite{Roul2015}. While both the $a$-plane and $m$-plane surfaces are non-polar, in our calculations we found that the $m$-plane $E_\mathrm{VBM}$ of AlN, GaN, and w-BN are within the $a$-plane $E_\mathrm{VBM}$ by only 0.2 eV. Therefore, we consider only the $a$-plane band alignments in this study, shown in \autoref{fig:band_alignment_no_SSE}c for AlN, \balnn, and GaN. All of the lowest formation energy $a$-plane slab models (Figures S13-S16) and their potential profiles (Figure S17) are available in the Supplementary Information.

In \baln $a$-plane surfaces, based on the average $E_{\mathrm{VBM}}$, we find a strong trend where $E_{\mathrm{VBM}}$ lowers roughly linearly with $x$. The w-BN $E_\mathrm{VBM}$ is roughly $0.7$ eV below that of AlN. Additionally, the increasing bandgap of \baln with $x$ means $E_\mathrm{CBM}$ stays relatively constant with $x$. We observe generally type-I band alignments of \baln with AlN, where the AlN band energies are located within the \baln energy gap. 

In general, the average $a$-plane $E_\mathrm{VBM}$ and $E_\mathrm{CBM}$ are systematically higher in energy ($\sim 2.3$ eV) than those of the $c$-plane interfaces. This trend has been observed experimentally in GaN\cite{Mishra2015} and theoretically in InN\cite{Belabbes2011} nonpolar and polar surfaces. We attribute this effect to differences in electronic structure from metallic and semiconducting surface states in polar and nonpolar surfaces, respectively\cite{Belabbes2011}. Interestingly, this leads to negative electron affinity (EA) for $x> 0.167$ surfaces, while the EA are all positive for $c$-plane \baln surfaces. Thus, high-$x$ $a$-plane \baln surfaces provide a route for spontaneous electron emitters that can be interfaced with other III-nitrides.

However, the band alignments with \baln depend strongly on the surface stoichiometry for the $a$-plane surfaces. There are huge standard deviations ($\leq1.4$ eV) in the predicted $E_{\mathrm{VBM}}$ and $E_\mathrm{CBM}$ of \baln structures for the same $x$. The standard deviations are higher in \baln slabs with lower tetrahedrality\cite{Milne2023} and peak at $x=0.6$. Therefore, this can be largely attributed to higher deviations in the tetrahedral bonds, particularly near the surface. Bond distortion induces dipoles near the surface due to differences in Al-N and B-N bond lengths as well as rotations in the bond angles. These effects dramatically influence the predicted band alignments, leading to high standard deviations.

Additionally, this effect is also related to the Al/B stoichiometry near the surface. We find that higher presences of surface B atoms at the surface lower the band alignment energies, similar to our observations for the $c$-plane slabs. Therefore, there is a wide range of possibilities for band alignment tuning in \baln heterostructures by controlling the interfacial stoichiometry. We find that both type-I and type-II band alignments in \balnn/AlN and \balnn/GaN nonpolar heterostructures could be realized.

There are very few VBO measurements involving non-polar AlN, GaN, or BN interfaces. In one $a$-plane VBO measurement for AlN/GaN, the VBO in $a$-plane AlN/GaN (GaN/AlN) was measured to be $0.73 \pm 0.16$ eV ($1.33 \pm 0.16$ eV)\cite{Sang2014}, which is much higher than our calculations--we calculate a VBO of 0.20 eV for the AlN/GaN heterojunction in the $a$-plane surface. 
This suggests that experimental effects such as intermixing in the AlN/GaN non-polar interfaces could influence the determination of their band alignments strongly, in line with findings from previous theoretical work\cite{Kaewmeechai2020b}.

\begin{figure}[h!]
 \includegraphics[width=0.85\columnwidth]{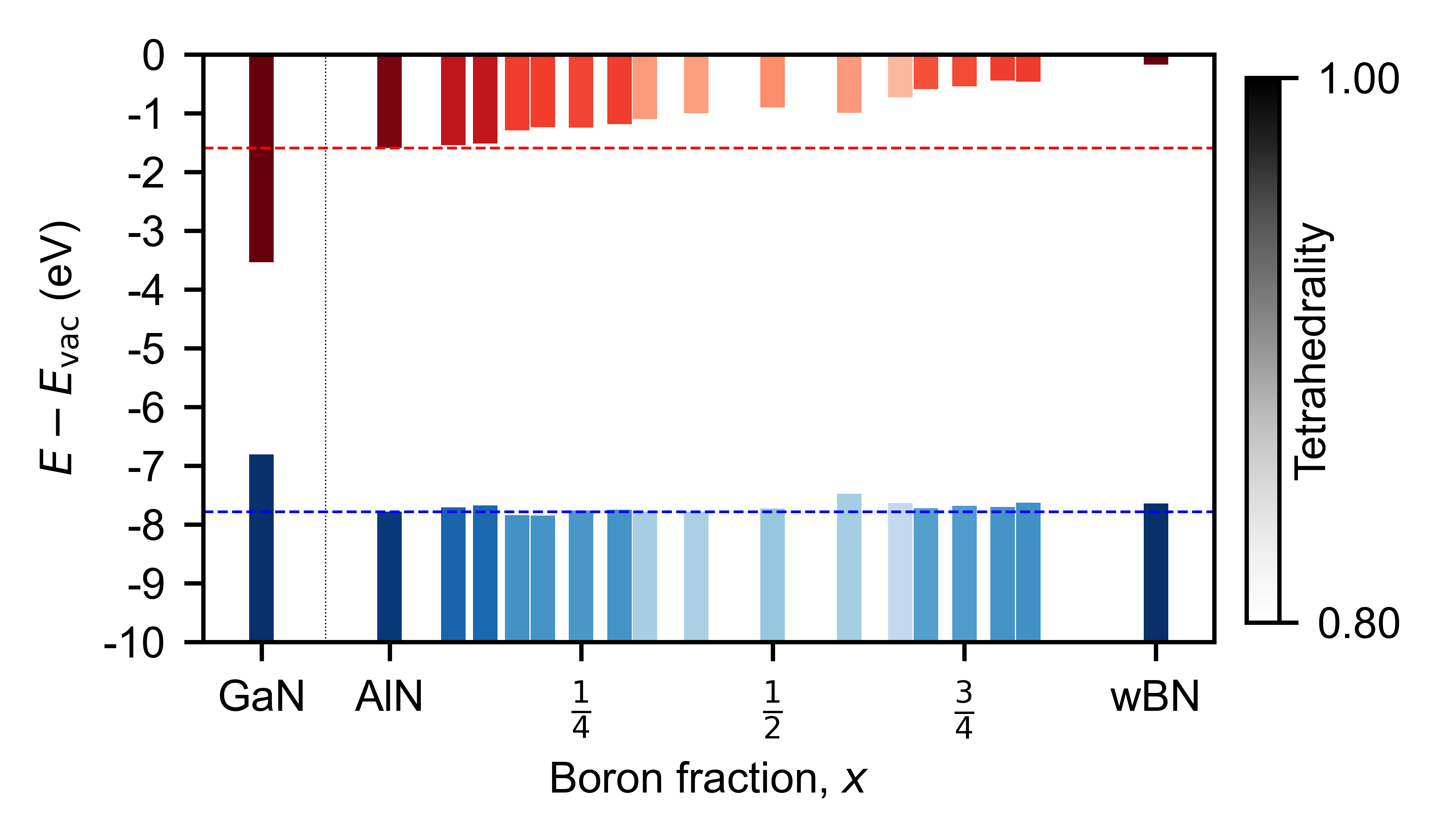}
 \caption{Natural band alignments for GaN and \baln calculated using the SSE approach, where $E_\mathrm{VBM}$ and $E_\mathrm{CBM}$ are represented by blue and red columns respectively, shaded according to the average tetrahedrality in the structure. Blue and red dashed lines show the AlN $E_\mathrm{VBM}$ and $E_\mathrm{CBM}$ respectively. Dashed gray lines separate the GaN and \baln results. The opacity of the columns is proportional to the average tetrahedrality of the corresponding bulk structure.}
 \label{fig:band_alignment_SSE}
\end{figure}

To compare the different approaches for calculating band alignments, we compute the natural band alignments for GaN and \baln similar to the solid-state energy (SSE) approach as implemented to study \baln by Ota \emph{et al.}\cite{Ota2022}, shown in \autoref{fig:band_alignment_SSE}. In the SSE model, $E_\mathrm{VBM}$ and $E_\mathrm{CBM}$ are calculated using only the bandgap and the SSE. The SSE depends only on the stoichiometry and a set of empirical parameters for each element, making this an extremely efficient method. For the cations, the SSE was taken to be the average electron affinities of 3, 5, and 5 main group compounds for B, Al, and Ga, respectively. For N, the SSE was taken to be the average ionization potential for the 5 main group compounds of N. Thus we obtained the SSE values as 2.18, 3.14, 3.82, and 7.0 eV, with standard deviations of 1.74, 0.80, 0.47, and 1.09 eV for B, Al, Ga and N, respectively\cite{Pelatt2019, Ota2022}. The $E_\mathrm{VBM}$ and $E_\mathrm{CBM}$ can then be computed using $E_{\mathrm{VBM}} = -\xi_\mathrm{M} - 0.5 E_g$ and $E_{\mathrm{CBM}} = -\xi_\mathrm{M} + 0.5 E_g$, where $E_g$ is the bandgap, and $\xi_\mathrm{M}$ is the geometric mean of the SSE of the constituent atoms\cite{Ota2020, Ota2022}.

\autoref{fig:band_alignment_SSE} shows that the $E_\mathrm{VBM}$ and $E_\mathrm{CBM}$ predicted from the SSE approach are systematically lower than those found in $c$-plane calculations but higher than those found in $a$-plane calculations. Since the SSE approach calculates $E_{\mathrm{VBM}}$ and $E_{\mathrm{CBM}}$ without consideration of interfacial effects (i.e. natural band alignments), the polar and non-polar contributions are effectively averaged.

In \balnn, $E_{\mathrm{VBM}}$ is nearly constant with $x$, while $E_{\mathrm{CBM}}$ increases linearly with $x$ due to the increasing bandgap. However, the EA values are always positive, with w-BN having the lowest predicted EA of 0.17 eV. Similar trends can be found in the natural band alignment of AlN, GaN, and w-BN\cite{Dreyer2014}.

Using SSE values, the AlN/GaN VBO is 0.98 eV, which is within the large range of experimental measurements. In comparison with the \balnn/GaN system however, we find that the B$_{0.125}$Al$_{874}$N/GaN VBO of 0.87 eV from SSE agrees much more poorly with the $c$-plane VBO measurement of $0.2 \pm 0.2$ for B$_{0.14}$Al$_{0.86}$N/GaN\cite{Sun2018b} than our $c$-plane calculation. There is better agreement with the B$_{0.13}$Al$_{0.87}$N/GaN measured VBO of $1.0 \pm 0.1$ by Rather \emph{et al.}, although the B$_{0.13}$Al$_{0.87}$N/AlN VBO of $0.5\pm0.1$ eV compares less favorably to the value of $-0.11$ eV for B$_{0.125}$Al$_{0.875}$N/AlN found using the SSE approach\cite{Rather2023}.

Therefore, while the SSE approach provides a computationally inexpensive method for estimating band alignments in bulk materials, it is limited by the available data. Surface-dependent effects are important to consider for reliable band alignment prediction\cite{Kaewmeechai2020b} and can change $E_{\mathrm{VBM}}$ by up to several eV in this system. In particular, this is important for B-containing compounds. The SSE for B is calculated using a low number of data points (3) with a high standard deviation ($1.74$ eV), which helps to explain the large discrepancies between measured and calculated VBO in \balnn/GaN and \balnn/AlN systems, as well as the non-linearities observed in our surface plane-dependent calculations.

In conclusion, we have employed DFT methods to study the band alignments of AlN, GaN, and \baln for $c$-plane and $a$-plane slabs, finding generally good agreement with the available experimental data. We use $GW_0$ methods for accurate bandgap prediction without the use of any empirical parameters. These calculations constitute the first surface-dependent band alignment predictions for \baln and GaN. For the polar $c$-planes, generally, low VBO are found for low-$x$ \balnn/AlN heterostructures, but there are nonlinearities at intermediate $x$ ($0.333 \leq x \leq 0.667$) \balnn. The nonlinearities are correlated with lower tetrahedrality in the bulk \baln structures. $E_\mathrm{CBM}$ of nearly all \baln structures is higher than that of AlN due to the increased bandgap, resulting in type-II (staggered) band alignments in \balnn/AlN heterostructures. In the non-polar $a$-plane surfaces, the average \baln $E_\mathrm{VBM}$ decreases with $x$ with a near constant $E_\mathrm{CBM}$, indicating that $a$-plane \baln heterojunctions have potential electron-conductive layer applications. However, there is a strong dependence on the distortion of Al-N and B-N tetrahedral bonds near the surface. We provide a comparison to band alignments calculated using the SSE approach, demonstrating that the surface effects are important to consider. These results mark a huge first step toward establishing the performance of \balnn/AlN and \balnn/GaN heterojunction-based devices. 

\begin{acknowledgments}
This work was supported by ULTRA, an Energy Frontier Research Center funded by the U.S. Department of Energy (DOE), Office of Science, Basic Energy Sciences (BES), under Award No. DESC0021230. This research used computing resources from the San Diego Supercomputer Center under the NSF-XSEDE and NSF-ACCESS Award No. DMR150006, and the Research Computing facility at Arizona State University. 
\end{acknowledgments}

The data that support the findings of this study are available from the corresponding author upon reasonable request.

\bibliography{main}

\end{document}